\documentclass{ws-ijmpd}
\usepackage[super,compress]{cite}

\usepackage{epsfig}
\usepackage{bm}
\usepackage{latexsym}
\usepackage{url}
\usepackage{dcolumn}
\usepackage{color}
\usepackage[usenames,dvipsnames]{xcolor}
\usepackage{amsfonts,amssymb,amsmath}
\usepackage{graphicx,epsfig}
\usepackage{psfrag}
\usepackage{subfigure}
\usepackage{comment}
\usepackage[utf8]{inputenc}
\usepackage{amsmath}
\usepackage{hyperref}
\hypersetup{colorlinks=true}

\begin{document}

\title{A plausible solution to the problem of initial conditions for inflation
}

\author{Suratna Das}
% \email{suratna@iitk.ac.in}
 \address{Department of Physics, Indian Institute of Technology, Kanpur 208016, India \\
suratna@iitk.ac.in }
\author{Raghavan Rangarajan}%etc.
%\email{raghavan@ahduni.edu.in}
\address{School of Arts and Sciences, Ahmedabad University, Navrangpura, Ahmedabad 380009, India \\
raghavan@ahduni.edu.in}

\maketitle
\begin{history}
\received{Day Month Year}
\revised{Day Month Year}
\end{history}

\begin{abstract}
We propose yet another solution to the initial condition problem of
inflation associated with homogeneity beyond the horizon at the onset
of inflation, in cases where inflation is preceded by a radiation era.
One may argue that causality will allow for smoothness over the causal
horizon scale $H^{-1}$, but for thermal inflationary scenarios, the
background inflaton field will only be correlated over the thermal
correlation length $\zeta\sim T^{-1}$ which is much smaller than
$H^{-1}$. We argue, with examples, that if the number of relativistic
degrees of freedom in the pre-inflationary era is very large $(>10^4)$
then the thermal correlation length can be of the order of the causal
horizon size alleviating the initial conditions problem of inflation.
%\Keywords{Inflation \and Initial Condition Problems}
\end{abstract}

\keywords{Inflationary Cosmology, Initial Conditions Problems}

\ccode{PACS numbers:}

%\maketitle

%-----------------------------------------------------------------------%
%%%%%%%%%%%%%%%%%%%%%%%%%%%%%

Despite the extraordinary success of the inflationary paradigm in
explaining cosmological observations, the initial conditions for
inflation are still being debated in modern cosmology
\cite{Ijjas:2013vea, Carrasco:2015rva, Brandenberger:2016uzh,
  Linde:2017pwt, Bagchi:2017stz, Mishra:2018dtg}.  The issue of
fine-tuned initial conditions claimed to be required for the onset of
inflation is quite multifaceted and poses concerns for the paradigm's
`real success'. Some of such issues are transition from kination to
inflation (the problem of entering an inflationary phase from a
kinetic dominated phase), the `unlikeliness problem' of inflation (the
issue recently raised in \cite{Ijjas:2013vea} concerning the form of
the inflationary potential favoured by the recent data), the issue of
initial large anisotropy and inhomogeneity, to name a few (for a more
recent review, see \cite{Chowdhury:2019otk}) . Here we focus on the
issue of initial inhomogeneity i.e. the requirement that the inflaton
should be homogeneous over a sufficiently large region for a causal
sized patch of the Universe to enter a period of inflation.

%=========================================================================================================================

{\it The issue of initial inhomogeneity}: Inflation was originally
proposed to tackle the severe fine-tuning problems of the Big Bang
Universe, namely the `horizon problem' and the `flatness problem'.
The success of inflation is only to be taken seriously if inflation
commences without any requirement of very special (fine-tuned) initial
conditions.  However, for one to apply the Robertson-Walker metric to
a horizon-sized patch at the beginning of inflation, and to argue that
inhomogeneities are small and irrelevant to the equation of motion of
the homogeneous component of the inflaton field, it is argued that
there was a patch somewhat bigger than the Hubble size which was
relatively homogeneous at the beginning of inflation
\cite{Goldwirth:1991rj, Vachaspati:1998dy}.  This requirement of an
assumption of homogeneity on superhorizon scales to start inflation
certainly seems incompatible with the motivation of the inflationary
paradigm to solve the horizon problem.

An analytical approach which is often used in the literature to
investigate such an initial condition problem is known as the
``effective-density approximation" (see \cite{Goldwirth:1989pr} and
for a recent review Sec.~IIIB of \cite{Chowdhury:2019otk}), where
inhomogeneties much smaller than the Hubble scale ($k\ll aH$)
(inhomogeneities much larger than the Hubble scale contribute to the
background evolution) can be analysed to show that even when the
initial energy density in the inhomogeneities ($\rho_{\delta\phi}$)
dominates over the energy density in the homogeneous field
($\rho_\phi$), the patch eventually enters an inflationary phase as
$\rho_{\delta\phi}$ decays as radiation in the course of time.
In early studies \cite{Albrecht:1984qt,
  Albrecht:1985yf,Albrecht:1986pi} it was shown analytically and
numerically for new inflation models that if the inflaton field is
inhomogeneous it will become homogeneous if the Hubble damping force
exceeds the effect of the $V^\prime(\phi)$ term in the field equation
of motion.  This was studied for the Coleman-Weinberg potential and
double well potential and with and without an initial thermal state.
Similar studies for chaotic inflation were carried out in
\cite{Kung:1989xz,Brandenberger:1990wu} and it was
found that chaotic inflation is an attractor in initial conditions
space -- one still obtains inflation despite the presence of density
inhomogeneities in the initial state provided that a single long
wavelength mode is excited. These studies did not consider
perturbations in the metric
%Such an approach also fails to analyse the effects of
and did not consider
inhomogeneities with  wavelengths comparable to the Hubble scale $(k\sim aH)$.
%, which renders the approach incomplete \cite{Chowdhury:2019otk}.
 Metric perturbations along with scalar field
  perturbations, in an expanding Universe, were considered in
  \cite{Feldman:1989hh} for chaotic inflationary models. In
\cite{Muller:1989rp} it was shown that power law inflation with an
exponential potential is a local attractor for inhomogeneous
cosmological models.  To consider perturbations of order the Hubble
size and to include perturbations in the metric a numerical relativity
approach was adopted in
\cite{Goldwirth:1989pr,Goldwirth:1989vz,Goldwirth:1990pm} for 1
dimension models.  It was argued there in that homogeneity over a few
initial horizons is needed for inflation.  Early numerical relativity
analysis in 3 dimensions for the inflationary scenario with
inhomogeneities were carried out in Ref. \cite{KurkiSuonio:1993fg}.
It has been recently shown in \cite{East:2015ggf} through numerical
simulations of full nonlinear Einstein equations that inflation does
commence even if the initial gradients are much larger than the
potential energy (with fluctuation length scales as large as the
Hubble scale). This result holds true if the inflaton fluctuations are
contained within the slow-rolling flat part of the inflaton potential,
and breaks down in cases where the gradients within the Hubble scale
feel the steep slope of the potential which results in break-down of
the slow-roll.  The conclusions of the 3 dimensional numerical
solutions is that for large field inflation the inflationary slow-roll
trajectory is a local attractor; however it is not a global attractor.
Embedding large field inflation in an ultraviolet complete quantum
field theory such as string theory has been difficult and if one is
forced to consider small field inflation then the initial conditions
problem will be severe \cite{Brandenberger:2016uzh}. In
  a numerical study \cite{Clough:2016ymm} with Einstein gravity in 3+1
  dimensions the authors compare the robustness of small and large
  field inflation against the initial inhomogeneities to show that
  small field inflation can fail, unlike the large field models, in
  the presence of even subdominant gradient energies.  In
\cite{Chowdhury:2019otk} it is also reiterated that it may be more
difficult to address the initial conditions problem for small field
inflation compared to plateau-like and large field inflation, and that
the initial conditions problem remains an important issue for
inflation.

It has also been proposed in \cite{Guth:2013sya} that if one assumes a
domination of negative spatial curvature $(k<0)$ in the patch which
will inflate (locally the patch will resemble an open FRW Universe)
then the curvature term will scale just like the gradient term
$(1/a^{2}(t))$, and the scale factor will scale as $a(t)\sim t$, just
like the Hubble radius, and this will resolve the above mentioned
conundrum.

The situation is somewhat different if one assumes the inflaton to be
in thermal equilibrium in a radiation dominated Universe till
inflation commences.  One may naively argue that if the inflaton is in
thermal equilibrium till the beginning of inflation then the inflaton
field will be homogeneous over the Hubble size at the beginning of
inflation.  Now for the inflaton field $\phi$ one may write
$\phi(x)=\phi_0(x) + \phi_Q(x)$,
%\begin{equation}
%\phi(x)=\phi_0(x) + \phi_Q(x),
%\end{equation}
where $\phi_0$ represents the background field whose slowly varying
potential energy drives inflation.  If the inflaton quanta are
scattering off each other in thermal equilibrium then there can be a
homogeneous distribution of the inflatons over some (thermal) length
scale, but that does not necessarily apply to the background field.
However, if the initial position of the background field is obtained
via a thermal phase transition at the beginning of inflation then one
may argue that the background inflaton field too can be correlated
over the (thermal) length scale

But, as was argued in \cite{Mazenko:1985pu}, when the inflaton field
is in thermal equilibrium at a temperature $T$, then there is no
statistical correlation beyond a scale $\zeta\sim T^{-1}$. Thus one
can set the scale of homogeneity for a thermalized Universe to be set
by the thermal correlation length $\zeta\sim T^{-1}$ and not by the
Hubble size $H^{-1}$ \cite{Mazenko:1985pu, Bagchi:2017stz}.  Therefore
the scale of homogeneity of the correlated inflaton field (and quanta)
will be given by $\zeta$, and
\begin{eqnarray}
\frac{\zeta}{H^{-1}}=\sqrt{\frac{\pi^2 g}{90}}\frac{T}{M_{\rm P}}\,,
\label{correlation1}
\end{eqnarray}
where $g$ is the effective number of relativistic degrees of freedom
of the radiation fluid, and we have used the condition that the energy
density of the inflaton field (and not the inflaton quanta) that
dominates the Universe equals that of the radiation at the onset of
inflation.  Then, for example, at the GUT scale
\begin{eqnarray}
\frac{\zeta_{\rm GUT}}{H^{-1}_{\rm GUT}}=\sqrt{\frac{\pi^2 g}{90}}\frac{T_{\rm GUT}}{M_{\rm P}}\sim \frac{T_{\rm GUT}}{M_{\rm P}}\sim 10^{-2},
\label{correlationGUT}
\end{eqnarray}
with $g\sim 100$ considering GUT particles as the constituents of the
radiation bath.  Thus one Hubble volume at the GUT scale contains
$10^6$ uncorrelated volumes.  The problem is worse at scales less than
the GUT scale.  If one considers a scenario such as natural inflation
\cite{Freese:1990rb, Adams:1992bn} in which the background inflaton
field correlation is determined from a phase transition at a
temperature scale $f$ higher than the inflaton scale $\Lambda$ then
the correlation length of the background inflaton field at the time of
inflation, presuming a thermal Universe from the scale $f$ to
$\Lambda$, will be $f^{-1} a(t_\Lambda)/a(t_f) = f^{-1}
T_f/T_\Lambda=\Lambda^{-1}$ which is again the thermal correlation
length at the beginning of inflation, and is less than the Hubble
length.

A few solutions to the initial conditions problem in an inflationary
Universe with a pre-inflationary radiation dominated era have been
proposed in the literature.  One may consider a compact (open or flat)
radiation dominated Universe, which has the topology of a torus (as
generally occurs in String Theories), and an effect called ``chaotic
mixing'' can lead to a rapid homogenization of the Universe
\cite{Linde:2004nz, Cornish:1996st}.  The problem with such a scenario
is that the inflaton must be at the very flat part of the inflaton
potential at the end of the pre-inflationary radiation era in order to
start inflation. Such a possibility is not easy to achieve in new
inflation or hilltop inflation scenarios, because of the shape of the
potential in these scenarios, while it is easy to achieve in
plateau-like potentials. Also, a long enough period of inflation is
required to make all these early Universe topological effects
unobservable \cite{Carrasco:2015rva}.  More recently it has been shown
in \cite{Bastero-Gil:2016mrl} that the thermalisation of the inflaton
field modes, via fluctuation-dissipation dynamics, for plateau-like
potentials can also lead to an earlier phase of thermal inflation
which causes a suppression of the effect of inhomogeneities on
inflation.

Another attempt has been made to resolve the initial conditions
problem in the case of natural inflation by modifying the inflaton
dynamics in the pre-inflationary radiation era \cite{Bagchi:2017stz}.
In this scenario one initially has a small region, smaller than the
Hubble size, over which the field is high on the potential and
homogenesous.  Due to the reaction-diffusion equation, the inflaton
changes very slowly in this region thereby maintaining a large
potential energy.  Then the expansion of the Universe leads to vacuum
energy domination inside a Hubble domain with a homogeneous field
throughout the domain triggering inflation.  This scenario also
presumes a radiation bath during the pre-inflationary era.
Consequences of such modified field dynamics on the observables are
yet to be analyzed.

 In \cite{Fairbairn:2017krt} it has been shown that the effects of the
 Gibbons-Hawking radiation before and during inflation for an inflaton
 coupled to a large number of spectator fields
 $(\mathcal{O}(10^{5.5}))$ can lead to a slow phase transition that
 helps to localise the inflaton at the top of its potential and create
 a scenario where the field is correlated across many horizons.

%\begin{comment}

In the rest of this Letter we will present another solution to the
initial conditions problem.  We will consider the scenario of a
pre-inflationary radiation era where the inflaton goes through a
thermal phase transition at the onset of inflation, so that the
correlation length for domains of the background inflaton field is
determined by the temperature at the beginning of inflation.  We will
argue that if the number of relativistic degrees of freedom in the
thermal bath is very large it can make the thermal correlation length
of the background inflaton field of order the Hubble size at the onset
of inflation.

%\end{comment}

{\it Another plausible solution}: We can see from
Eq.~(\ref{correlationGUT}) that the presence of a large number of
relativistic degrees of freedom (say $g\sim10^4$) during the
pre-inflationary radiation era gives the correlation length at the GUT
scale $\zeta_{\rm GUT}\sim H_{\rm GUT}^{-1}$.  For fixed $H_{\rm GUT}$
or energy density $\rho_{\rm GUT}$, increasing $g$ decreases $T_{\rm
  GUT}$ and hence increases $\zeta_{\rm GUT}$.  (If the value of $g$
is large enough that $T^{-1}$ is greater than $H^{-1}$, then the
length scale over which homogeneity can be assumed would be the Hubble
size and not $T^{-1}$.)  For inflationary scales lower than the GUT
scale one would need $g$ to be larger than $10^4$ to obtain a
homogeneous inflationary patch of the size of the Hubble length.  We
now present two scenarios where such large values of $g$ in the
pre-inflationary radiation bath with a thermalised inflaton are
incorporated for other reasons.  Then our arguments above provide
another mechanism for addressing the initial conditions problem for
these scenarios.

I. {\it Pre-inflationary radiation era with a thermalized inflaton in
  `just enough' inflation}:

A thermalized inflaton during the pre-inflationary radiation era was
first considered in \cite{Bhattacharya:2005wn} where the thermalized
inflaton decoupled from the thermal bath during the radiation era,
leading to a frozen thermal Bose-Einstein distribution of the inflaton
field.  It is natural to consider that the inflaton might have been in
thermal equilibrium sometime during the radiation era at some very
early time, as the inflaton must have some coupling with the other
fields in order to reheat the Universe at the end of inflation.  This
scenario of the thermalized decoupled inflaton was revisited in
\cite{Das:2014ffa} where along with the frozen distribution of the
inflaton field, the effect of the pre-inflationary radiation era on
the inflaton mode functions was also considered, and the best fit
value of the duration of inflation in such a case is obtained to be
less than one e-folding more than the `just enough' case. The best fit
curve also sets an upper bound on the comoving temperature of the
inflaton quanta as $\mathcal{T}<10^{-4}$ Mpc$^{-1}$. (The comoving
temperature in \cite{Das:2014ffa,Das:2015ywa} is represented as $T$.)

It was later recognised in \cite{Das:2015ywa} that the upper bound on
the comoving temperature $\mathcal{T}$ of the inflaton and the minimal
number of extra e-foldings $\delta N$ over the `just enough'
inflationary case preferred by the data make the scenario of
pre-inflationary radiation era with a thermalized decoupled inflaton
incompatible with observations.  Considering that in such a scenario
the radiation energy density would be equal to the inflaton potential
energy density at the onset of inflation, it can be shown that
\begin{eqnarray}
\frac{H_I}{M_{\rm P}}=\left(\frac{90}{\pi^2 g}\right)^{1/2}\left(\frac{H_0}{\mathcal{T}}\right)^2 e^{-2\delta N}
> 1,
\label{HIrelation}
\end{eqnarray}
where 
$H_0$ is the Hubble parameter
 at the present time and $g\sim 100$. 
But the observational bound on the inflationary scale yields \cite{Ade:2015lrj}
\begin{eqnarray}
\frac{H_I}{M_{\rm P}}\lesssim 1.56\times 10^{-5},
\label{HIbound}
\end{eqnarray}
for $r\lesssim 0.07$.  Alternatively for $\mathcal{T}<10^{-4}$
Mpc$^{-1}$, and $\delta N\approx 0.05$ and $H_0=(4400\,{\rm
  Mpc})^{-1}$ one gets from the equality in Eq.~(\ref{HIrelation}) and
Eq.~(\ref{HIbound}) that $g\gtrsim10^{12}$, for the scenario to be
compatible with the data.

The above scenario of `just enough' inflation is for an inflaton that
was once thermal but decoupled early in the history of the Universe.
If, however, the scenario of `just enough' inflation with $g\gtrsim
10^{12}$ to be compatible with observations is extended to have the
inflaton in thermal equilibrium till the beginning of inflation, then
the corresponding thermally correlated region with a homogeneous
inflaton field will be as large as the Hubble volume, thereby
addressing the initial conditions problem.

It may be noted that there are good theoretical arguments for
considering scenarios of `just enough' inflation.  It has been argued
in \cite{Freivogel:2005vv} that the number of e-foldings of inflation
should not be much larger than the minimum required to satisfy
observations, by considering anthropic bounds on the spatial curvature
of the Universe and including certain statistical arguments on the
parameters of inflation.  In \cite{Gibbons:2006pa} it has been argued
that the probability for $N$ e-foldings of inflation is suppressed by
a factor of $\exp(-3N)$, by invoking a certain natural canonical
measure on the space of all classical Universes.  Moreover, more
inflation means that the inflaton potential is required to stay flat
for a larger range of field values and this can then require
super-Planckian excursions.

II. {\it Warm inflation:} The warm inflationary scenario
\cite{Berera:1995ie}, where the is creating a
radiation bath during inflation, is a well known alternative scenario
to the standard supercooled inflationary scenario. Such an
inflationary scenario starts with a radiation dominated era, enters an
inflationary regime, and then transits into a radiation dominated
epoch, with or without invoking any reheating phase in between.  In
\cite{Berera:2000xz} it is argued that the smoothness requirement on
the initial inflationary patch is on scales smaller than $H^{-1}$.
However this is true for $\Gamma>H$, where $\Gamma$ is the dissipation
rate of the inflaton field, while $\Gamma<H$ (weak dissipation) warm
inflation scenarios have also been widely considered in the
literature.

Viable models of warm inflation require $10^6$ or $10^4$ additional
fields to satisfy warm inflation requirements, particularly to
maintain $T > H$ during inflation
\cite{Berera:1998px,Bastero-Gil:2015nja}.
Then, our discussion above implies that this can provide for a
background inflaton field correlated at the beginning of inflation
over a Hubble volume, if the initial background inflaton field value
is obtained from a thermal phase transition.

{\it In search of a large number of relativistic degrees of freedom}:
It is now legitimate to ask whether one can indeed devise a particle
physics model where such a large number of relativistic degrees of
freedom can be realised.  A particle physics model inspired by string
theory with $N=1$ supersymmetry was constructed in
\cite{Berera:1998px, Berera:1998cq} where the inflaton is coupled to
$N_M(>10^4)$ number of chiral superfields representing the string
modes. Each of the superfields will have its anti-chiral
superfield. In the Lagrangian the inflaton field has couplings to
$N_M\times N_\chi$ scalars and $N_M\times N_\psi$ fermions, where
setting $N_\psi=N_\chi/4$ helps cancel the radiatively generated
vacuum energy correction in the effective potential. Assuming all the
particles are in thermal equilibrium, it yields the number of
relativistic degrees of freedom to be of order $10^6$ if one takes
$N_M\sim \mathcal{O}(10^5)$.  A higher $N_M$ would result in a higher
number of relativistic degrees of freedom.  Thus having a large number
of relativistic degrees of freedom in the early Universe is feasible,
and it points towards more exotic physics taking place at the GUT
scale and beyond.

{\it Discussion and Conclusion}: In models of inflation where the
inflaton goes through a thermal phase transition the background
inflaton field can be correlated on scales of order the thermal
correlation length $\zeta\sim T^{-1}$, which is typically much smaller
than $H^{-1}$ for standard cosmological scenarios with a thermal bath
of $O(100)$ relativistic degrees of freedom when inflation commences.
We have argued that if the number of relativistic degrees of freedom
is of the order of $10^4$ for, say, GUT scale inflation then the
thermal correlation length is of the size of the Hubble scale.  We
have also shared two inflationary scenarios where one desires to have
such a large number of degrees of freedom for other reasons.  In the
first case of a thermalized inflaton in a `just enough' inflation
scenario, the presence of a large number of relativistic degrees of
freedom in the pre-inflationary radiation bath is required to ensure
the scenario is in accordance with observations.  In the context of
warm inflation such large numbers are required to keep the temperature
during warm inflation higher than the Hubble parameter.

One may ask if this scenario of the inflaton going through a thermal
phase transition at the onset of inflation is feasible.  If the
inflaton was in thermal equilibrium in the very early Universe and
subsequently decoupled there will be a frozen thermal-like
distribution of inflaton particles in the Universe at the beginning of
inflation.  This scenario has been studied in
Refs. \cite{Bhattacharya:2005wn,Das:2014ffa,Das:2015ywa}.  The frozen
distribution of inflaton particles will contribute to the effective
potential of the inflaton field and, for an appropriate form of the
potential, one can have a thermal-like phase transition at $T$, the
notional temperature of the frozen distribution equal to the original
decoupled temperature from early time scaled by the scale factor,
decreases with time.  This will give a correlation length of the
inflaton field of order $T^{-1}$.  Furthermore, we have also
considered warm inflation in which the inflaton can have a thermal
effective potential.  For an appropriate choice of the potential one
can obtain a thermal phase transition at the onset of inflation.

The above proposal partially resolves the initial conditions problem
associated with homogeneity on superhorizon scales at the onset of
inflation, for inflationary models preceded by a thermal
  radiation era.  What we have proposed leads to correlations of the
background inflaton field at least up to the Hubble size which alleviates the
initial conditions problem substantially.  
In the absence of our proposal the correlation length
for models of inflation with a thermal history  
will be much smaller than the Hubble size. 

\section*{Acknowledgments}
%\begin{acknowledgements}
We would like to thank Hiranmaya Mishra, Ajit Srivastava and Rudnei
O. Ramos for valuable discussions. We would also like to thank the
referee for highlighting certain important references. The work of
S.D. is supported by Department of Science and Technology, Government
of India under the Grant Agreement number IFA13-PH-77 (INSPIRE Faculty
Award).
 %\end{acknowledgements}

\label{Bibliography}
\bibliographystyle{ws-ijmpd}
\bibliography{ini-cond}

\end{document}